\pdfoutput=1

\documentclass{article}

\usepackage[final]{neurips_2022}

\usepackage[utf8]{inputenc}
\usepackage[T1]{fontenc}
\usepackage{hyperref}
\usepackage{url}
\usepackage{booktabs}
\usepackage{amsfonts}
\usepackage{nicefrac}
\usepackage{microtype}
\usepackage{xcolor}
\usepackage{amsmath}
\usepackage{hyperref}
\usepackage{blindtext}
\usepackage{graphicx}

\makeatletter\let\NAT@force@numbers\relax\makeatother

\title{Best Prompts for Text-to-Image Models\\and How to Find Them}

\author{%
  Nikita Pavlichenko \\
  Toloka \\
  11000 Belgrade, Serbia \\
  \texttt{pavlichenko@toloka.ai} \\
  \And
  Dmitry Ustalov \\
  Toloka \\
  11000 Belgrade, Serbia \\
  \texttt{dustalov@toloka.ai} \\
}

\begin{document}

\maketitle

\begin{abstract}
Advancements in text-guided diffusion models have allowed for the creation of visually appealing images similar to those created by professional artists. The effectiveness of these models depends on the composition of the textual description, known as the \emph{prompt}, and its accompanying keywords. Evaluating aesthetics computationally is difficult, so human input is necessary to determine the ideal prompt formulation and keyword combination. In this study, we propose a human-in-the-loop method for discovering the most effective combination of prompt keywords using a genetic algorithm. Our approach demonstrates how this can lead to an improvement in the visual appeal of images generated from the same description.
\end{abstract}

\section{Introduction}

Recent progress in computer vision and natural language processing has enabled a wide range of possible applications to generative models. One of the most promising applications is text-guided image generation (text-to-image models). Solutions like DALL-E~2~\citep{ramesh2022} and Stable Diffusion~\citep{rombach2021highresolution} use the recent advances in joint image and text embedding learning (CLIP~\citep{pmlr-v139-radford21a}) and diffusion models~\citep{pmlr-v37-sohl-dickstein15} to produce photo-realistic and aesthetically-appealing images based on a textual description.

However, in order to ensure the high quality of generated images, these models need a proper \emph{prompt engineering} \citep{Liu_2022} to specify the exact result expected from the generative model. In particular, it became a common practice to add special phrases (\emph{keywords}) before or after the image description, such as ``trending on artstation,'' ``highly detailed,'' etc. Developing such prompts requires human intuition, and the resulting prompts often look arbitrary. Another problem is the lack of evaluation tools, so practically, it means that the user subjective judges the quality of a prompt by a single generation or on a single task. Also, there is currently no available analysis on how different keywords affect the final quality of generations and which ones allow to achieve the best images aesthetically.

In this work, we want to bridge this gap by proposing an approach for a large-scale human evaluation of prompt templates using crowd workers. We apply our method to find a set of keywords for Stable Diffusion that produces the most aesthetically appealing images. Our contributions can be summarized as follows:
\begin{itemize}\itemsep0em
  \item We introduce a method for evaluating the quality of generations produced by different prompt templates.
  \item We propose a set of keywords for Stable Diffusion and show that it improves the aesthetics of the images.
  \item We release all the data and code that allow to reproduce our results and build solutions on top of them, such as finding even better keywords and finding them for other models.
\end{itemize}

\section{\label{sec:prompts}Prompts and How to Evaluate Them}

\begin{figure}[t]
  \centering
  \includegraphics[width=\textwidth]{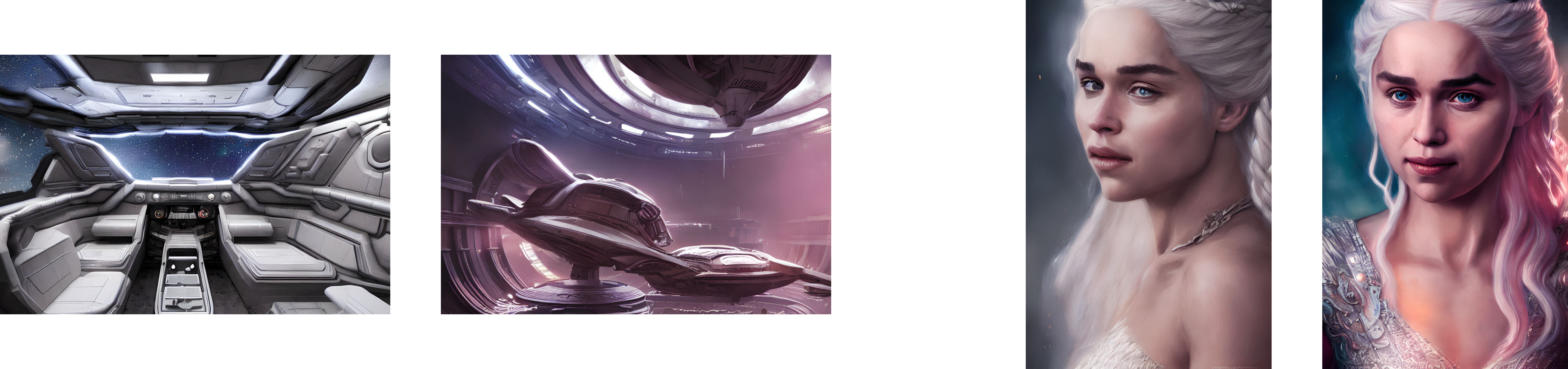}
  \caption{\label{fig:example_images}Comparison of the keyword sets. Left: no keywords vs. our approach. Right: 15 most popular keywords vs. our approach. Images are cherry-picked.}
\end{figure}

Consider a standard setup for generative models with text inputs. A model gets as an input a natural language text called \emph{prompt} and outputs a text completion in the case of the text-to-text generation or an image in the case of text-to-image generation. Since specifying the additional information increases the quality of the output images \citep{Liu_2022}, it is common to put specific keywords before and after the image description:
\begin{equation*}
    \text{prompt} = [\text{kw}_1,\ldots,\text{kw}_{m-1}]\;[\text{description}]\;[\text{kw}_m,\ldots,\text{kw}_{n}].
\end{equation*}

Consider a real-world example when a user wants to generate an image of a cat using a text-to-image model.\footnote{\url{https://lexica.art/prompt/28f5c644-9310-4870-949b-38281328ffd0}} Instead of passing a straightforward prompt
\emph{a cat}, they use a specific prompt template, such as \emph{Highly detailed painting of a calico cat, cinematic lighting, dramatic atmosphere, by dustin nguyen, akihiko yoshida, greg tocchini, greg rutkowski, cliff chiang, 4k resolution, luminous grassy background}. In this example, the \textbf{description} is \emph{painting of a calico cat} and the \textbf{keywords} are \emph{highly detailed, cinematic lighting, dramatic atmosphere, by dustin nguyen, akihiko yoshida, greg tocchini, greg rutkowski, cliff chiang, 4k resolution, luminous grassy background}.

Since aesthetics are difficult to evaluate computationally, we propose a human-in-the-loop method for evaluating the keyword sets. Our method takes as an input a set of textual image descriptions $\mathcal{D}$, a set of all possible keywords $\mathcal{K}$, and a set of the keyword set candidates $\mathcal{S}$ and outputs a list of keyword sets $s_i \subseteq \mathcal{K}, s_i \in \mathcal{S}$ in the increasing order of their aesthetic appeal to humans. Since it is challenging for annotators to directly assign scores for images or rank them, we run pairwise comparisons of images generated from a single description but with different keyword sets and then infer the ranking from the annotation results. Our algorithm can be described as follows:
\begin{enumerate}
  \item For each pair of a description $d_i \in \mathcal{D}$ and a keyword set $s_j \in \mathcal{S}$, generate four images $I_{ij} = \{I_{ij_1}, \ldots, I_{ij_4}\}$.
  \item For each image description $d_i \in \mathcal{D}$, sample $nk\log_2(n)$ pairs of images $(I_{ij}, I_{ik})$ generated with different keyword sets, where $n$ is the number of keyword sets to compare, and $k$ is the number of redundant comparisons to get the sufficient number of comparisons \citep{pmlr-v70-maystre17a}.
  \item Run a pairwise comparison crowdsourcing task in which the workers are provided with a description and a pair of images, and they have to select the best image without knowing the keyword set.
  \item For each description $d_i \in \mathcal{D}$, aggregate the pairwise comparisons using the Bradley-Terry probabilistic ranking algorithm \citep{Bradley_1952}, recovering a list $r_i = s_1 \prec \cdots \prec s_n$ of keyword sets ordered by their visual appeal to humans.
  \item For each keyword set, compute the average rank in the lists recovered for the descriptions.
\end{enumerate}

As a result, we quantify the quality of a keyword set as its rank averaged per description.

\section{Iterative Estimation of the Best Keyword Set}

\begin{figure*}[t]
  \centering
  \includegraphics[width=1.0\textwidth]{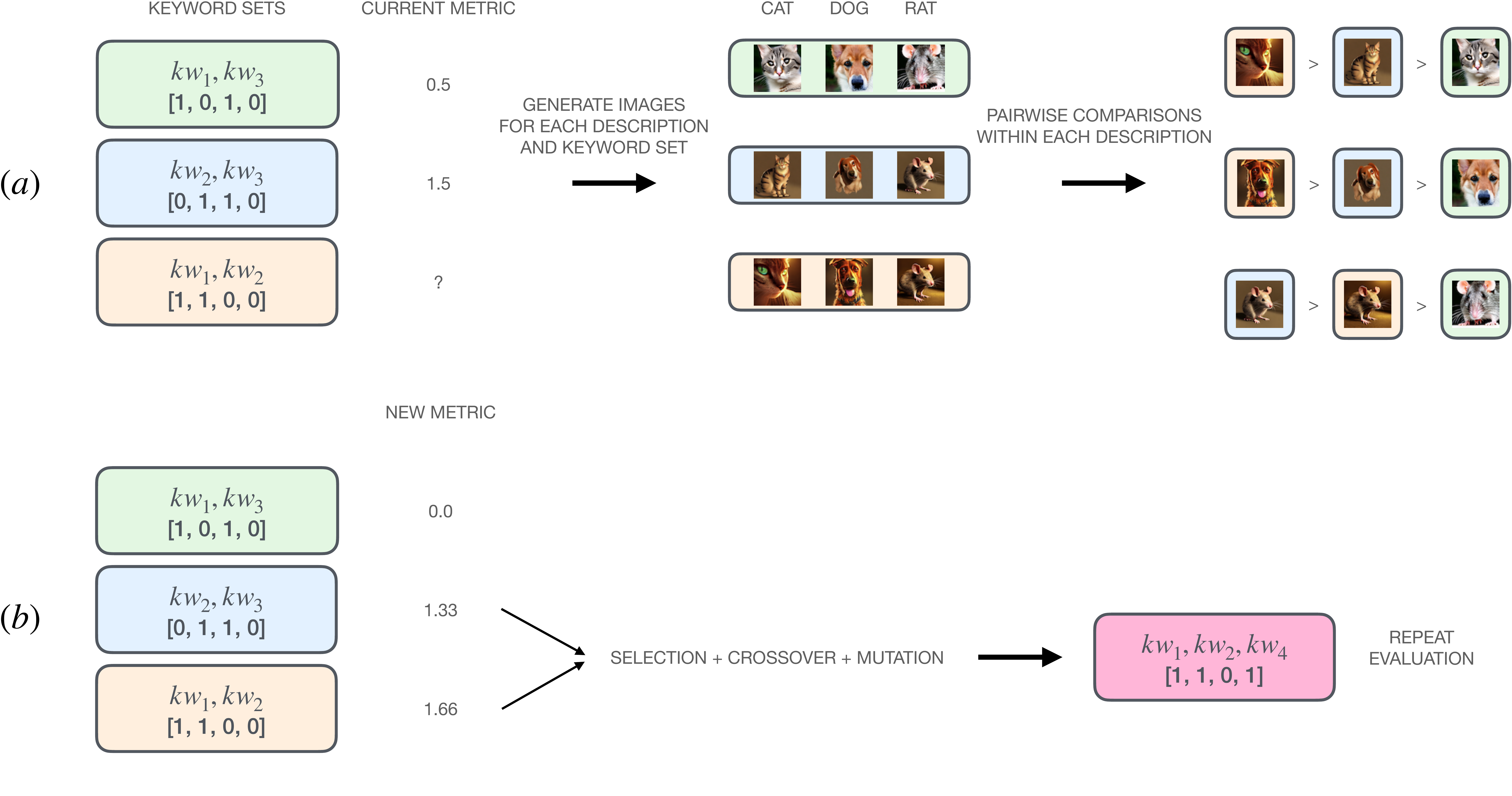}
  \caption{\label{fig:algorithm}A scheme of genetic optimization of keyword sets. (a) Evaluation of a new candidate keyword set: first, we generate images for all descriptions with a new keyword set; second, we run pairwise comparisons of generated images within each description between the previous and new keyword sets to obtain the ranking. The average rank of keyword sets is a quality metric. (b) We take two keyword sets with the highest rank and perform crossover and mutation to obtain a new candidate, which is then evaluated according to scheme (a). The process is repeated for the pre-determined number of iterations.}
\end{figure*}

One of the advantages of our approach is that the keywords can be evaluated iteratively. Once we have compared a number of keyword sets, we can request a small additional number of comparisons to evaluate the new set of keywords. This allows us to apply discrete optimization algorithms, such as a genetic algorithm, to retrieve from a large pool of keywords the most influential keywords.

Figure \ref{fig:algorithm} represents a scheme of our approach. We pick a set of keyword sets for initialization, rank the keywords using the approach in Section~\ref{sec:prompts}, and use it as an initial population for the genetic algorithm. Then we repeat the following steps multiple times to obtain the best-performing keyword set.
\begin{enumerate}
  \item Obtain the next candidate keyword set $s_j$ based on quality metrics of currently evaluated keyword sets using the genetic algorithm. We present the details on a particular variation of a genetic algorithm we use in Section \ref{sec:setup}.
  \item For each image description $d_i \in \mathcal{D}$, sample
  $k\left((n+1)\log_2(n + 1) - n\log_2 n\right)$ pairs $(I_{ik}, I_{ij})$ of images generated using keywords from the new candidate set and already evaluated keyword sets. We do this to sustain $kn\log_2 n$ comparisons in total.
  \item Evaluate the quality of the obtained keyword set (steps 3--5 in Section~\ref{sec:prompts}).
\end{enumerate}

\section{\label{sec:exp}Experiment}

We perform an empirical evaluation of the proposed prompt keyword optimization approach in a realistic scenario using the publicly available datasets.

\subsection{\label{sec:setup}Setup}

To construct a set of possible keywords, we have parsed the Stable Diffusion Discord\footnote{\url{https://discord.com/invite/stablediffusion}} and took the 100 most popular keywords. For image descriptions, we decided to choose prompts from six categories: \emph{portraits}, \emph{landscapes}, \emph{buildings}, \emph{interiors}, \emph{animals}, and \emph{other}. We took twelve prompts for each category from Reddit and \url{https://lexica.art/} and manually filtered them to obtain only raw descriptions without any keywords.

We use a simple genetic algorithm to find the optimal prompt keyword set. The algorithm was initialized with two keyword sets: one is an empty set, and another set contained the 15 most popular keywords that we retrieved before. We limited the maximum number of output keywords by 15 as otherwise, the resulting prompts became too long.

In order to evaluate the keyword sets, we generate four images for each prompt constructed by appending comma-separated keywords to the image description in alphabetical order. Each image was generated with the Stable Diffusion model \citep{rombach2021highresolution} with 50 diffusion steps and 7.5 classifier-free guidance scale using the DDIM scheduler~\citep{song2020denoising}. Then, we run crowdsourcing annotation on the Toloka crowdsourcing platform.\footnote{\url{https://toloka.ai/}} The crowd workers need to choose the most aesthetically-pleasing generated images in $3n \log_2 n$ pairs (we set $k = 3$ as we have a limited budget) for each image description, where $n$ is the number of currently tried keyword sets. Textual pseudographics of the annotation interface is shown in Figure~\ref{fig:interface}.

\begin{figure}
    \centering
    \begin{tabular}{lr}\toprule
    % \begin{tabular}{lr}\toprule
         \multicolumn{2}{c}{\textbf{Interior of an alien spaceship}} \\\midrule
         \texttt{Image L1} & \texttt{Image R1} \\
         \texttt{Image L2} & \texttt{Image R2} \\
         \texttt{Image L3} & \texttt{Image R3} \\
         \texttt{Image L4} & \texttt{Image R4} \\\midrule
         \multicolumn{2}{l}{Which set is better? \qquad $\square$ Left \enspace $\square$ Right} \\\bottomrule
    % \end{tabular}
    \end{tabular}
    \caption{Textual pseudographics of the annotation interface. A crowd worker sees two sets of four images generated for a single description but with different keyword sets (one on the left and one on the right) and needs to choose the more aesthetically-pleasing set of images.}
    \label{fig:interface}
\end{figure}

Since crowdsourcing tasks require careful quality control and our task involved gathering subjective opinions of humans, we followed the synthetic golden task production strategy proposed for the IMDB-WIKI-SbS dataset~\citep{imdb-wiki-sbs}. We randomly added comparisons against the images produced by a simpler model, DALL-E~Mini~\citep{Dayma_DALL_E_Mini_2021}. We assumed that DALL-E~Mini images are less appealing than the ones generated by Stable Diffusion, and choosing them was a mistake. Hence, we suspended the workers who demonstrated an accuracy lower than 80\% on these synthetic golden tasks.

After the annotation is completed, we run the Bradley-Terry~\citep{Bradley_1952} aggregation from the Crowd-Kit~\citep{CrowdKit} library for Python to obtain a ranked list of keyword sets for each image description. The final evaluation metric used in the genetic algorithm to produce the new candidate sets is the average rank of a keyword set (as described in Section~\ref{sec:prompts}). We use 60 image descriptions for optimization (ten from each category) and 12 for the validation of the optimization results.

For the keywords optimization, we use a genetic algorithm as follows. We parameterized every keyword set by a binary mask of length 100, indicating whether the keyword should be appended to the prompt. We initialized the algorithm with all zeros and the mask including the 15 most popular keywords. At the selection step, we took the two masks with the highest average rank. At the crossover step, we swapped a random segment of them. At the mutation step, we swapped bits of the resulting offsprings with
probability of 1\% to get the resulting candidates.

\subsection{Results}

\begin{table*}[t]
\centering
\caption{\label{tab:results}Average rank of the baseline keywords (top-15 most common on Stable Diffusion Discord) and the ones found by the genetic algorithm. Rank is averaged over 60 prompts on train and over 12 prompts on validation (val); maximal rank is 56.}
\resizebox{\textwidth}{!}{\begin{tabular}{rrrrcrrrr}\toprule
\multicolumn{4}{c}{\textbf{Train}} &  \phantom{abc} & \multicolumn{4}{c}{\textbf{Validation}} \\
\cmidrule{1-4} \cmidrule{6-9}
No Keywords & Top-15 & Best Train & Best Val & & No Keywords & Top-15 & Best Train & Best Val \\ \midrule
$3.5$ & $14.25$ & $\boldsymbol{43.60}$ & $39.32$ & & $5.42$ & $12.50$ & $38.00$ & $\boldsymbol{46.00}$ \\
\bottomrule
\end{tabular}}

\end{table*}

\begin{figure*}[t]
  \centering
  \includegraphics[width=0.8\textwidth]{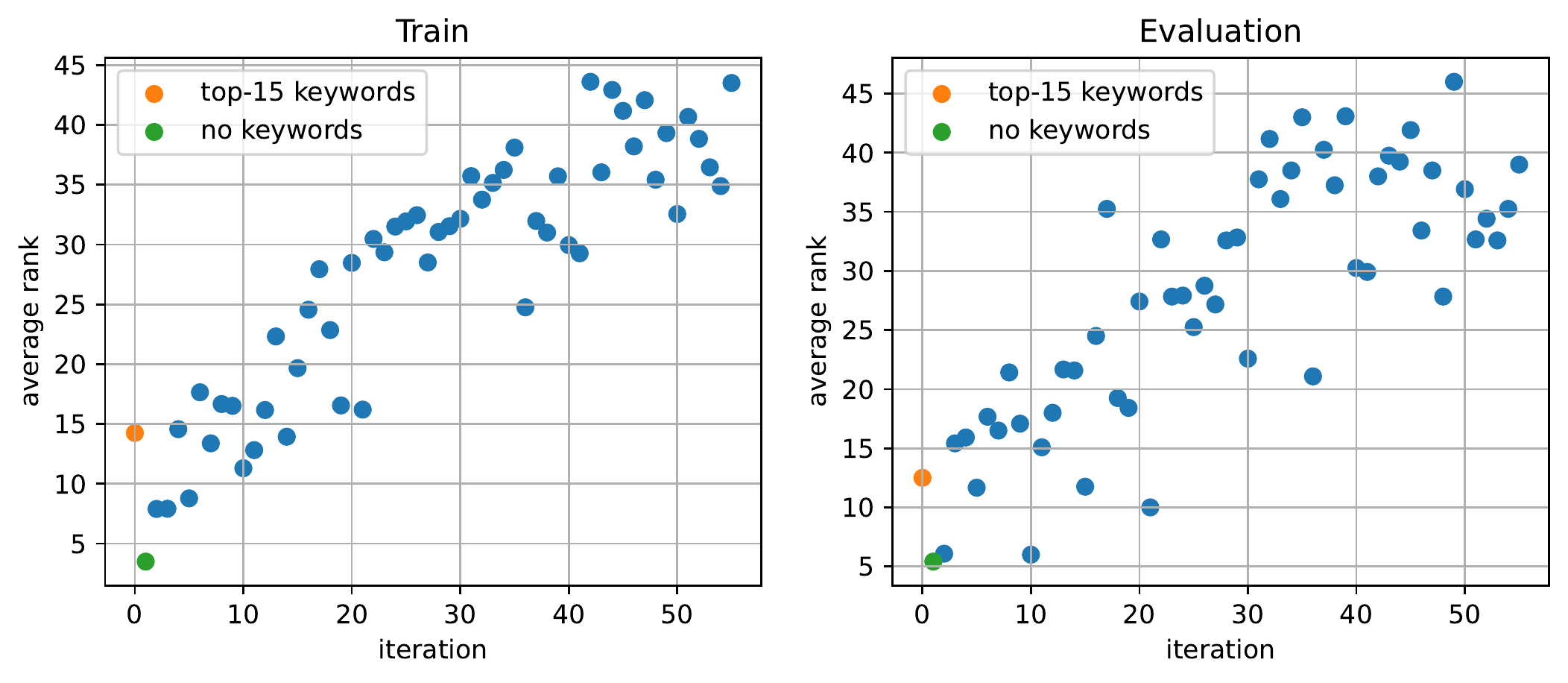}
  \caption{\label{fig:genetic_alg}Average ranks of keyword sets tried by the genetic algorithm. There are total 56 keyword sets, so the maximal average rank is 56.}
\end{figure*}

We ran the optimization for 56 iterations on 60 image descriptions since we have a fixed annotation budget. To ensure that our method did not overfit, we ran the evaluation on another 12 descriptions (validation). Figure~\ref{fig:genetic_alg} shows ranks of tried keyword sets. According to the evaluation results in Table~\ref{tab:results}, we found that our algorithm was able to find a significantly better set of keywords than the fifteen most popular ones (Top-15). Also, we see that any set of prompt keywords is significantly better than no keywords at all (No Keywords).

We see that most results hold on the validation set, too, but the metrics have more noise. Overall, the best set of keywords on the training set of 60 prompts is \emph{cinematic, colorful background, concept art, dramatic lighting, high detail, highly detailed, hyper realistic, intricate, intricate sharp details, octane render, smooth, studio lighting, trending on artstation}. An example of images generated with this keyword set is shown in Figure~\ref{fig:example_images}.

\subsection{Discussion}

We show that adding the prompt keywords significantly improves the quality of generated images. We also noticed that the most popular keywords do not result in the best-looking images. To estimate the importance of different keywords, we trained a random forest regressor~\citep{Breiman:01} on the sets of keywords and their metrics that is similar to W\&B Sweeps.\footnote{\url{https://docs.wandb.ai/guides/sweeps}} We found that the most important keywords, in reality, are different from the most widely used ones, such as ``trending on artstation.'' The most important keyword we found was ``colorful background.''

There are several limitations to our approach. We can not conclude that the found set of keywords is the best one since the genetic algorithm can easily fall into a local minimum. In our run, it tried only 56 keywords out of the 100 most popular ones. Also, our evaluation metrics are based on ranks, not absolute scores, so they are not sensitive enough to determine the convergence of the algorithm.

However, since we release all the comparisons, generated images, and code, it is possible for the community to improve on our results. For instance, one can run a genetic algorithm from a different initialization, for a larger number of iterations, or even with more sophisticated optimization methods. This can easily be done by comparing the new candidates with our images and adding these results to the dataset.

\section{Related Work}

The aesthetic quality evaluation is one of the developing topics in computer vision. There are several datasets and machine learning methods aiming at solving this problem~\citep{Sheng_2018,zhang2021image}. However, the available datasets contain human judgments on image aesthetics scaled from 1 to 5. Our experience shows that the pairwise comparisons that we used in this paper are a more robust approach as different humans perceive scales differently and subjectively. Also, they specify training a model to evaluate the aesthetics but not on the generative models. Large language models, such as GPT-3~\citep{GPT3}, have enabled a wide range of research tasks on prompt engineering~\citep{BetterFewShot,Liu2022PretrainPA,fantasticallyOrderedPrompts,CrossTaskGeneralization,HowContextAffects,PromptProgramming,RetrivePrompts}. Recent papers also discover the possibilities of prompt engineering for text-to-image models and confirm that prompts benefit from the added keywords~\citep{Liu_2022}. To the best of our knowledge, we are the first to apply it to find the best keywords.

% https://www.inovex.de/de/blog/prompt-engineering-guide/

\section{Conclusion}

We presented an approach for evaluating the aesthetic quality of images produced by text-to-image models with different prompt keywords. We applied this method to find the best keyword set for Stable Diffusion and showed that these keywords produce better results than the most popular keywords used by the community. Despite the fact that our work focuses on the evaluation of keywords for text-to-image models, it is not limited by this problem and can be applied for an arbitrary prompt template evaluation, for example, in the text-to-text setting. This is a direction for our future work. Last but not least, we would like to encourage the community to continue our experiment and find better keyword sets using our open-source code and data.\footnote{\url{https://github.com/toloka/BestPrompts}}

% \clearpage

\bibliographystyle{plain}
\bibliography{prompts}

\begin{thebibliography}{10}

\bibitem{Bradley_1952}
Ralph~Allan Bradley and Milton~E. Terry.
\newblock {Rank Analysis of Incomplete Block Designs: I. The Method of Paired
  Comparisons}.
\newblock {\em Biometrika}, 39(3/4):324--345, 1952.

\bibitem{Breiman:01}
Leo Breiman.
\newblock {Random Forests}.
\newblock {\em Machine Learning}, 45(1):5--32, 2001.

\bibitem{GPT3}
Tom Brown et~al.
\newblock {Language Models are Few-Shot Learners}.
\newblock In {\em Advances in Neural Information Processing Systems 33},
  NeurIPS~2020, pages 1877--1901, Montr\'{e}al, QC, Canada, 2020. Curran
  Associates, Inc.

\bibitem{Dayma_DALL_E_Mini_2021}
Boris Dayma, Suraj Patil, Pedro Cuenca, Khalid Saifullah, Tanishq Abraham, Phuc
  Le~Khac, Luke Melas, and Ritobrata Ghosh.
\newblock {DALL-E Mini}, 2021.

\bibitem{BetterFewShot}
Tianyu Gao, Adam Fisch, and Danqi Chen.
\newblock {Making Pre-trained Language Models Better Few-shot Learners}.
\newblock In {\em Proceedings of the 59th Annual Meeting of the Association for
  Computational Linguistics and the 11th International Joint Conference on
  Natural Language Processing (Volume~1: Long Papers)}, ACL-IJCNLP~2021, pages
  3816--3830, Online, 2021. Association for Computational Linguistics.

\bibitem{Liu2022PretrainPA}
Pengfei Liu, Weizhe Yuan, Jinlan Fu, Zhengbao Jiang, Hiroaki Hayashi, and
  Graham Neubig.
\newblock {Pre-Train, Prompt, and Predict: A Systematic Survey of Prompting
  Methods in Natural Language Processing}.
\newblock {\em ACM Computing Surveys}, 55(9), 2022.

\bibitem{Liu_2022}
Vivian Liu and Lydia~B Chilton.
\newblock {Design Guidelines for Prompt Engineering Text-to-Image Generative
  Models}.
\newblock In {\em Proceedings of the 2022 CHI Conference on Human Factors in
  Computing Systems}, CHI '22, New Orleans, LA, USA, 2022. Association for
  Computing Machinery.

\bibitem{fantasticallyOrderedPrompts}
Yao Lu, Max Bartolo, Alastair Moore, Sebastian Riedel, and Pontus Stenetorp.
\newblock {Fantastically Ordered Prompts and Where to Find Them: Overcoming
  Few-Shot Prompt Order Sensitivity}, 2021.
\newblock arXiv:2104.08786.

\bibitem{pmlr-v70-maystre17a}
Lucas Maystre and Matthias Grossglauser.
\newblock {Just Sort It! A Simple and Effective Approach to Active Preference
  Learning}.
\newblock In {\em Proceedings of the 34th International Conference on Machine
  Learning}, volume~70 of {\em ICML~2017}, pages 2344--2353, Sydney, NSW,
  Australia, 2017. PMLR.

\bibitem{CrossTaskGeneralization}
Swaroop Mishra, Daniel Khashabi, Chitta Baral, and Hannaneh Hajishirzi.
\newblock {Cross-Task Generalization via Natural Language Crowdsourcing
  Instructions}.
\newblock In {\em Proceedings of the 60th Annual Meeting of the Association for
  Computational Linguistics (Volume 1: Long Papers)}, ACL~2022, pages
  3470--3487, Dublin, Ireland, 2022. Association for Computational Linguistics.

\bibitem{imdb-wiki-sbs}
Nikita Pavlichenko and Dmitry Ustalov.
\newblock {IMDB-WIKI-SbS: An Evaluation Dataset for Crowdsourced Pairwise
  Comparisons}, 2021.
\newblock arXiv:2110.14990.

\bibitem{HowContextAffects}
Fabio Petroni, Patrick Lewis, Aleksandra Piktus, Tim Rocktäschel, Yuxiang Wu,
  Alexander~H. Miller, and Sebastian Riedel.
\newblock {How Context Affects Language Models' Factual Predictions}, 2020.
\newblock arXiv:2005.04611.

\bibitem{pmlr-v139-radford21a}
Alec Radford, Jong~Wook Kim, Chris Hallacy, Aditya Ramesh, Gabriel Goh,
  Sandhini Agarwal, Girish Sastry, Amanda Askell, Pamela Mishkin, Jack Clark,
  Gretchen Krueger, and Ilya Sutskever.
\newblock {Learning Transferable Visual Models From Natural Language
  Supervision}.
\newblock In {\em Proceedings of the 38th International Conference on Machine
  Learning}, volume 139 of {\em ICML~2021}, pages 8748--8763, Virtual Only,
  2021. PMLR.

\bibitem{ramesh2022}
Aditya Ramesh, Prafulla Dhariwal, Alex Nichol, Casey Chu, and Mark Chen.
\newblock {Hierarchical Text-Conditional Image Generation with CLIP Latents},
  2022.
\newblock arXiv:2204.06125.

\bibitem{PromptProgramming}
Laria Reynolds and Kyle McDonell.
\newblock {Prompt Programming for Large Language Models: Beyond the Few-Shot
  Paradigm}.
\newblock In {\em Extended Abstracts of the 2021 CHI Conference on Human
  Factors in Computing Systems}, CHI EA '21, Yokohama, Japan, 2021. Association
  for Computing Machinery.

\bibitem{rombach2021highresolution}
Robin Rombach, Andreas Blattmann, Dominik Lorenz, Patrick Esser, and Bj\"orn
  Ommer.
\newblock {High-Resolution Image Synthesis With Latent Diffusion Models}.
\newblock In {\em Proceedings of the IEEE/CVF Conference on Computer Vision and
  Pattern Recognition (CVPR)}, pages 10684--10695, New Orleans, LA, USA, 2022.
  IEEE.

\bibitem{RetrivePrompts}
Ohad Rubin, Jonathan Herzig, and Jonathan Berant.
\newblock {Learning To Retrieve Prompts for In-Context Learning}, 2022.

\bibitem{Sheng_2018}
Kekai Sheng, Weiming Dong, Chongyang Ma, Xing Mei, Feiyue Huang, and Bao-Gang
  Hu.
\newblock {Attention-Based Multi-Patch Aggregation for Image Aesthetic
  Assessment}.
\newblock In {\em Proceedings of the 26th ACM International Conference on
  Multimedia}, MM '18, pages 879--886, Seoul, Republic of Korea, 2018.
  Association for Computing Machinery.

\bibitem{pmlr-v37-sohl-dickstein15}
Jascha Sohl-Dickstein, Eric Weiss, Niru Maheswaranathan, and Surya Ganguli.
\newblock {Deep Unsupervised Learning using Nonequilibrium Thermodynamics}.
\newblock In {\em Proceedings of the 32nd International Conference on Machine
  Learning}, volume~37 of {\em ICML~2015}, pages 2256--2265, Lille, France,
  2015. PMLR.

\bibitem{song2020denoising}
Jiaming Song, Chenlin Meng, and Stefano Ermon.
\newblock {Denoising Diffusion Implicit Models}, 2020.
\newblock arXiv:2010.02502.

\bibitem{CrowdKit}
Dmitry Ustalov, Nikita Pavlichenko, and Boris Tseitlin.
\newblock {Learning from Crowds with Crowd-Kit}, 2023.

\bibitem{zhang2021image}
Bo~Zhang, Li~Niu, and Liqing Zhang.
\newblock {Image Composition Assessment with Saliency-augmented Multi-pattern
  Pooling}, 2021.
\newblock arXiv:2104.03133.

\end{thebibliography}

\clearpage

\appendix

\textbf{\LARGE Appendix}

\section{Keyword Selection}

To find the most popular prompt keywords, we parsed the Stable Diffusion Discord \texttt{gobot} channel, collected the prompts the users submitted, and counted the phrases separated by commas. Then, we took the 100 most popular keywords ordered by their appearances in prompts. This approach resulted in a small amount of common phrases that often appeared in prompts but could not be considered as keywords. We manually filtered the keyword list to exclude them. Table~\ref{tab:keywords} presents the final list.

\section{Image Descriptions}

Tables~\ref{tab:trainprompts} and~\ref{tab:valprompts} present image descriptions we collected from \url{https://lexica.art/} and \url{https://old.reddit.com/r/StableDiffusion/}.

\section{Annotation}

We ran our annotation on Toloka. In each human intelligence task, the worker sees an image description without prompt keywords, four images on the left and four images on the right. They had to choose the more appealing set of images---left or right. Figure~\ref{fig:interface} shows our task interface.

We used the following approach for worker selection. First, we required the interested workers to pass a qualification test. During the test, they had to correctly identify five sets of images generated by Stable Diffusion from five sets of images generated by DALL-E Mini on a single page. Those who passed the test were allowed to earn money by annotating pairs of image sets. During annotation, one of five task pages contained a similarly-designed golden task. Those who made at least one mistake on these golden tasks were disqualified from our task. We also controlled the time workers spent to complete the task by suspending those who completed the task page faster than in 15 seconds. As a result, we 12,724 workers annotated 597,830 pairs, and accuracy on golden tasks was 84\%.

\section{Keywords Optimization}

We used a simple genetic algorithm to optimize the keyword sets. We parameterized every keyword set by a binary mask of length 100 indicating whether the keyword should be appended to the prompt. We initialized the algorithm with all zeros and the mask including the 15 most popular keywords. At the selection step, we took the two masks with the highest average rank. At the crossover step, we swapped a random segment of them. At the mutation step, we swapped bits of the resulting offsprings with
probability of 1\% to get the resulting candidates. Figure~\ref{fig:genetic_alg} shows ranks of tried keyword sets.

\section{Keyword Importance}

\begin{figure}[ht]
  \centering
  \includegraphics[width=\textwidth]{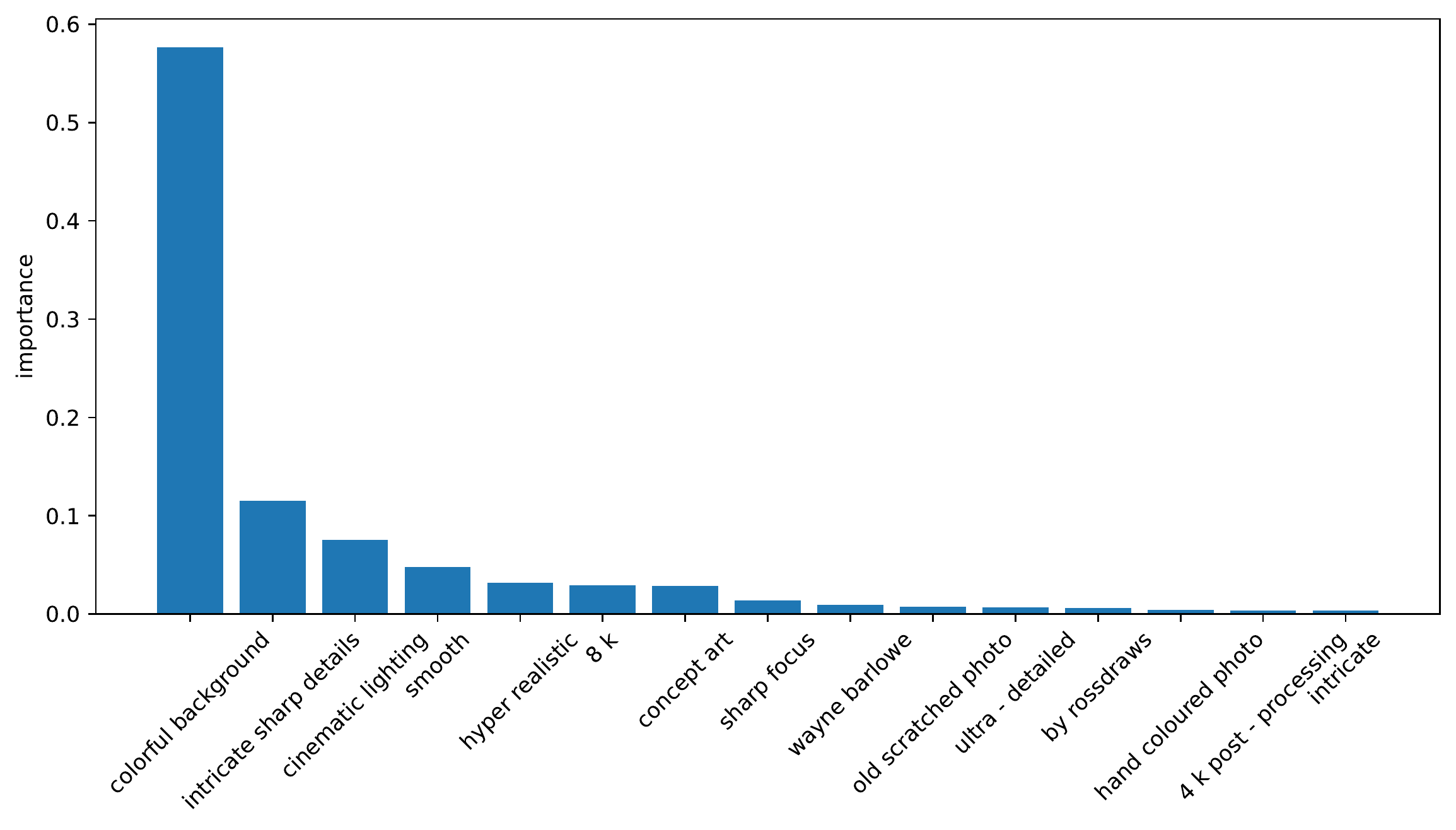}
  \caption{Importance of top-15 most important keywords.}
  \label{fig:importance}
\end{figure}

Figure~\ref{fig:importance} represents the importance of top-15 most important keywords estimated by training a random forest on a dataset containing keyword masks and their metric values. Note that higher importance does not always mean higher quality.

\begin{table}[ht]
\centering
\caption{\label{tab:keywords}Top-100 most common keywords and their appearances in \texttt{gobot} channel prompts.}
\resizebox{\textwidth}{!}{\begin{tabular}{lrlr}\toprule
\textbf{Keyword} & \textbf{\# of Appearances} & \textbf{Keyword} & \textbf{\# of Appearances} \\ \midrule
highly detailed & 6062 & insanely detailed & 527 \\
sharp focus & 3942 & wayne barlowe & 526 \\
concept art & 3539 & atmospheric & 515 \\
intricate & 3240 & by rossdraws & 504 \\
artstation & 2841 & hypermaximalist & 499 \\
digital painting & 2840 & pop surrealist & 498 \\
smooth & 2599 & boris vallejo & 489 \\
elegant & 2574 & by james jean & 478 \\
illustration & 2300 & frank franzzeta & 470 \\
cinematic lighting & 2152 & mcbess & 470 \\
octane render & 2090 & brosmind & 470 \\
trending on artstation & 2049 & steve simpson & 470 \\
8 k & 1864 & krenz cushart & 470 \\
dramatic lighting & 1322 & decadent & 468 \\
cinematic & 1253 & ilya kuvshinov & 463 \\
volumetric lighting & 1242 & by kyoto animation & 462 \\
greg rutkowski & 1118 & art by ruan jia and greg rutkowski & 461 \\
unreal engine & 1046 & mucha fantasy art artifacts & 460 \\
realistic & 1029 & hajime sorayama & 456 \\
4 k & 952 & aaron horkey & 456 \\
digital art & 942 & hyperrealistic & 452 \\
sharp & 941 & natural raw unreal tpose & 448 \\
unreal engine 5 & 879 & akihiko yoshida & 444 \\
pulp fiction & 875 & by greg rutkowski & 438 \\
focus & 792 & ultra realistic & 435 \\
hyper realistic & 779 & cosmic horror & 416 \\
colorful background & 745 & ultra detailed & 415 \\
vray & 726 & high detail & 414 \\
qled & 720 & 8k & 386 \\
finely detailed features & 710 & studio ghibli & 385 \\
detailed & 678 & ray tracing & 382 \\
perfect art & 627 & colorfully & 372 \\
trending on pixiv fanbox & 627 & photo realism & 368 \\
beautiful & 621 & matte & 361 \\
ominous & 614 & intricate sharp details & 335 \\
artgerm & 608 & dynamic compositiom & 321 \\
peter mohrbacher & 605 & volumetric light & 312 \\
fantasy intricate elegant & 599 & colorful & 310 \\
studio lighting & 599 & photorealism & 308 \\
craig mullins & 592 & ultra - detailed & 308 \\
photorealistic & 581 & hand coloured photo & 306 \\
digital airbrush & 570 & high definition & 303 \\
gaston bussiere & 561 & concept art artgerm & 298 \\
hyper realism & 555 & natural lighting & 297 \\
intricate details & 553 & collodion wet paint photo & 296 \\
sakimi chan & 546 & 4 k post - processing & 291 \\
studio quality & 545 & oil painting & 290 \\
magical illustration & 540 & photoreal & 289 \\
ornate & 540 & old scratched photo & 286 \\
matte painting & 535 & cgsociety & 283 \\
\bottomrule
\end{tabular}}
\end{table}

\begin{table}[ht]
\centering
\caption{\label{tab:trainprompts}Image descriptions used for training, their categories and orientations of the generated images.}
\resizebox{\textwidth}{!}{\begin{tabular}{p{14cm}rr}\toprule
\textbf{Image Description} & \textbf{Type} & \textbf{Orientation} \\ \midrule
A potrait of a space fanstasy cat & animals & portrait \\
An interstellar cat in a spacesuit & animals & square \\
wolf portrait, ferns, butterflies & animals & portrait \\
portrait photo of an armored demonic undead deer with antlers, in a magical forest looking at the camera & animals & album \\
Whale spaceship flying near a red dwarf star & animals & square \\
A portrait of a monstrous frog covered in blue flames & animals & portrait \\
The Highland Cow is a beautiful animal & animals & album \\
Vicious dog with three heads, glowing eyes and matted fur & animals & portrait \\
A golden tiger resting, dragon body & animals & portrait \\
sleeping cute baby turtle, under the sea & animals & album \\
Futuristic city center with 890j maglev train in background & buildings & album \\
painting of pripyat & buildings & square \\
Post apocalyptic shopping center, raining, building, avenue & buildings & album \\
Priests gathering at aztec pyramid in jungle & buildings & album \\
Photograph. Mordor photo. Manhattan photo & buildings & portrait \\
tokyo city market & buildings & portrait \\
Mars landscape futurstic city & buildings & square \\
X-Wing over Manhattan & buildings & album \\
steampunk city levitating above a large ocean & buildings & album \\
Dream fantasy in little european town & buildings & album \\
Vampires fighting in a party in the interior of gothic dark castle, red pool fountain, louis xv furniture & interior & square \\
Interior of an alien spaceship & interior & square \\
Halo 3 interiors & interior & square \\
A vast indoor growing operation on the edge of space, in a massive cavernous iron city & interior & album \\
Steampunk greenhouse interior & interior & album \\
Fallout interior render & interior & square \\
Computer repair. Woman building a dieselpunk computer. Glowing screens. Huge dieselpunk computer & interior & album \\
painting of a vast gothic library & interior & square \\
A Dark, Spooky and gloomy Haunted kitchen with lot of dried fruits and dried vegetables & interior & portrait \\
A Photo of Astronomers studying the night sky with a telescope inside Observatory & interior & album \\
silk road lanscape, rocket ship, space station & landscape & album \\
gigantic paleolothic torus made of stone with carvings of shamanic robotic electronics and circuitry, in a mediterranean lanscape, inside a valley overlooking the sea & landscape & square \\
night, the ocean, the milk way galaxy & landscape & album \\
Winterfell walls gate, lanscape & landscape & album \\
the river of time glowing in the dark & landscape & portrait \\
Beautiful meadow at sunrise, thin morning fog hovering close to the ground & landscape & album \\
a beach full of trash and dead animals, whales, fish & landscape & square \\
a comfortable survival shelter made out of a container home with an attached garden and a small tent extension on the side, exterior walls are made of transparent material allowing light to pass through, Yosemite national park green meadows with beautiful big redwood trees on the edge, Mountains in the background and a creek running calmly through the meadow, Blue hour and a visible milkyway in the sky & landscape & album \\
A mountain in the shape of wolf dental arch & landscape & portrait \\
Cabela's beautiful comfortable modular insulated wall kit - house all weather family dwelling tent house, person in foreground, mountainous forested wilderness open fields & landscape & album \\
Steampunk helmet mask robot & other & portrait \\
heaven made of fruit basket & other & square \\
An isolated apple tree & other & portrait \\
torus brain in edgy darkiron camel & other & portrait \\
Wrc rally car stylized & other & album \\
American phone booth with antenna in the woods & other & portrait \\
Blue flame captured in a bottle & other & square \\
depiction of the beginning of the universe inside a snow globe & other & square \\
A human skull floating in deep dark murky water & other & portrait \\
A Photograph of Cumulus Clouds emerging from a teacup & other & square \\
a portrait of a mafia boss in a golden suit & portrait & square \\
A portrait of a rough male farmer in world war 2, 1 9 4 0 setting & portrait & portrait \\
Portrait of a blue genasi tempest priest & portrait & portrait \\
Portrait of beautiful angel & portrait & album \\
Arab man light beard, curly hair, swordsman & portrait & portrait \\
Blonde-haired beautiful Warrior Queen, in fantasy armor, with Iron crown, cross symbolism, with a fit body, dark forest background & portrait & portrait \\
spacer woman, with Symmetric features, curly(changed to taste through gens) hair with realistic proportions, wearing rugged and torn workers clothes & portrait & square \\
rapunzel, wedding dress & portrait & square \\
Portrait of a knight, holding a sword, victorian & portrait & portrait \\
princess peach in the mushroom kingdom & portrait & square \\
\bottomrule
\end{tabular}}
\end{table}

\begin{table}[ht]\centering
\caption{\label{tab:valprompts}Image descriptions used for validation, their categories and orientations of the generated images.}
\resizebox{\textwidth}{!}{\begin{tabular}{lrr}\toprule
\textbf{Image Description} & \textbf{Type} & \textbf{Orientation} \\ \midrule
East - european shepard dog, portrait & animals & square \\
A painting of a horse in the middle of a field of flowers & animals & album \\
Medieval gothic city with castle on top of the hill & buildings & portrait \\
London in 2 0 5 0 & buildings & album \\
An empty science research laboratory & interior & album \\
Hogwarts great hall art & interior & album \\
A painting of a valley with black tree stumps and broken stone. scorched earth, sunset & landscape & square \\
Epic mountain view surrounded by lake & landscape & portrait \\
Floating glass sphere filled with a raging storm & other & square \\
Portrait shot of cybertronic airplane in a scenic dystopian environment & other & portrait \\
portrait of gabriel knight, from sierra adventure game & portrait & portrait \\
A portrait painting of daenerys targaryen queen & portrait & portrait \\
\bottomrule
\end{tabular}}
\end{table}

\begin{figure}[ht]
  \centering
  \includegraphics[width=\textwidth]{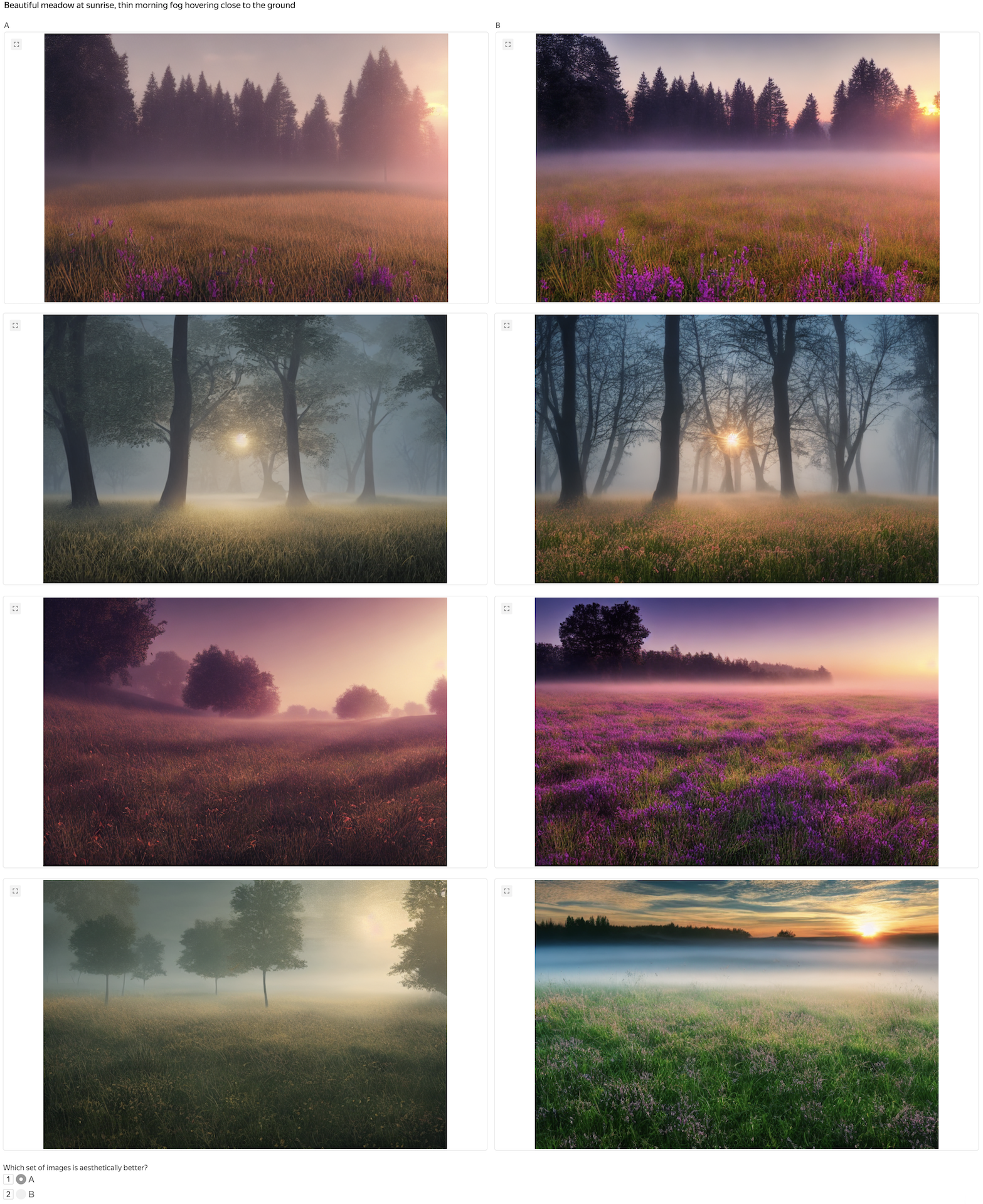}
  \caption{\label{fig:interface}Our annotation task interface.}
\end{figure}

\begin{figure}[ht]
  \centering
  \includegraphics[width=\textwidth]{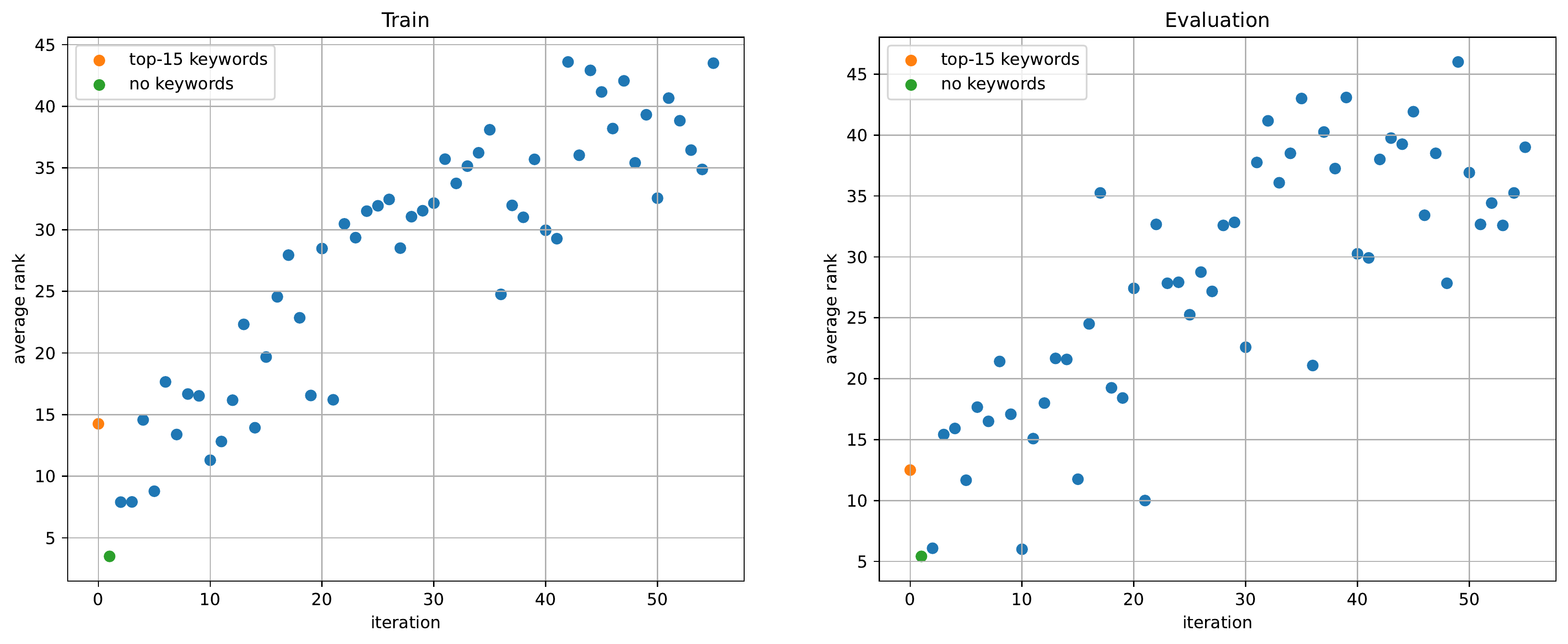}
  \caption{\label{fig:genetic_alg}Average ranks of keyword sets tried by the genetic algorithm. There were total 56 keyword sets, so the metric values are limited by 56.}
\end{figure}

\end{document}